\documentclass{article}

 \usepackage[dblblindworkshop,nonatbib,final]{neurips_2025}

\usepackage[utf8]{inputenc} %
\usepackage[T1]{fontenc}    %
\usepackage{hyperref}       %
\usepackage{url}            %
\usepackage{booktabs}       %
\usepackage{amsfonts}       %
\usepackage{nicefrac}       %
\usepackage{microtype}      %
\usepackage{xcolor}         %
\usepackage{amsmath}
\usepackage{cleveref}
\usepackage{tcolorbox}
\usepackage{subcaption}

 \usepackage[
  backend=biber,
  style=numeric,   %
  natbib=true,        %
  maxnames=6,         %
  minnames=6          %
]{biblatex}

\tcbuselibrary{listings,breakable}

\newcommand{\papername}[0]{AsymPuzl~}
\newcommand{\agenta}[0]{Alice~}
\newcommand{\agentb}[0]{Bob~}

\title{AsymPuzl: An Asymmetric Puzzle for multi-agent cooperation}
\workshoptitle{Multi-Turn Interactions in Large Language Models}

\author{%
    Xavier Cadet \thanks{\texttt{xavier.fjf.cadet@dartmouth.edu}}\\
    Dartmouth College \\
    Hanover, NH 03755
    \And
    Edward Koh \\
    Dartmouth College\\
    Hanover, NH 03755 \\
    \And
    Peter Chin \\
    Dartmouth College \\
    Hanover, NH 03755}

\begin{document}

\maketitle

\begin{abstract}
Large Language Model (LLM) agents are increasingly studied in multi-turn, multi-agent scenarios, yet most existing setups emphasize open-ended role-play rather than controlled evaluation.
We introduce AsymPuzl, a minimal but expressive two-agent puzzle environment designed to isolate communication under information asymmetry.
Each agent observes complementary but incomplete views of a symbolic puzzle and must exchange messages to solve it cooperatively.
Using a diverse set of current-generation and open-source LLMs, we show that (i) strong models such as GPT-5 and Claude-4.0 reliably converge across puzzle sizes on the solution by sharing complete information in two turns, (ii) weaker models often ignore partner messages or over-correct their hypotheses, and (iii) feedback design is non-trivial: simple self-feedback improves success rates, while detailed joint feedback can hurt performance.
These findings show that even in simple cooperative tasks, LLM communication strategies diverge and depend on the granularity of feedback signals.
AsymPuzl thus provides a testbed for probing the limits of multi-turn cooperation and opens avenues for studying coordination mechanisms.
\end{abstract}

\section{Introduction} \label{section:introduction}
Autonomous agents using Language Models (LLMs) offer promising opportunities for problem-solving \cite{bubeckSparksArtificialGeneral2023,brownLanguageModelsAre2020}.
Nonetheless, real-world problems often require multi-turn and collaboration under partial information.
In many tasks, agents often have access to complementary yet potentially incomplete information, as seen in distributed decision-making or human-ai cooperation.
Information asymmetry is a common phenomenon in practice and requires effective communication to address the issue.
Current Multi-Agent LLM studies emphasize \textit{role-play} and \textit{open-ended dialogue} but lack controlled testbeds for evaluating communication strategies.
Moreover, existing Large Language Model Multi-Agent Systems (LLM-MAS) rarely allow systematic manipulation of task difficulty.
As such, we are left with the question: \textit{How do LLM agents adapt their communication under asymmetric information.}
We introduce Asympuzl, a minimal yet expressive environment where two agents see complementary partial views of a puzzle and must exchange messages to solve it. We can adjust the difficulty of the task by varying the puzzle size and providing feedback on their hypotheses.
Our \textbf{contributions} are: i) the \textbf{\papername environment}, a testbed for two agent puzzle solving under information asymmetry, and ii) an \textbf{empirical analysis of feedback granularity and communication strategies} using this environment, showing that while feedback can be helpful, it can be lead to reduced performance if not carefully designed, and that while complete information sharing is possible in our experiments, most agents do not default to this strategy.

\section{Related Work} \label{section:related_work}
\paragraph{Puzzle-based reasoning:} Puzzle solving is a popular method to evaluate the reasoning capacities of LLMs \cite{tyagiStepbyStepReasoningSolve2024,dziriFaithFateLimits2023,liAssessingLogicalPuzzle2024,renVGRPBenchVisualGrid2025}.  
\cite{badolaMultiTurnPuzzlesEvaluating2025} compares LLMs over multi-turn puzzles with a single agent and shows that errors often come from poor instruction following. ZebraLogic \cite{linZebraLogicScalingLimits2025} isolates the reasoning limits of single LLMs on logical puzzles, demonstrating that as puzzle complexity increases performance of agents drop and increasing model size yields limited improvement. In comparison, this work isolates the communication and coordination in a multi-agent scenario when no single agent holds complete information (sufficient to solve the puzzle alone).
\paragraph{LLM-MAS:} Interest in coordinating multiple LLM agents has been increasing \cite{liCAMELCommunicativeAgents2023, eriskenMAEBEMultiAgentEmergent2025,wuAutoGenEnablingNextgen2024,liangEncouragingDivergentThinking2024} alongside concerns relating to emergent behavior during multi-agent interactions and model alignment. iAgents \cite{liuAutonomousAgentsCollaborative2024} focused on large-scale role-play social networks with information asymmetry. In comparison, our environment is deliberately minimal, designed to reduce the effect of confounding factors and provide fine grained insights into how LLMs adapt communication protocols.

\section{The \papername Environment} \label{section:methodology}
\begin{figure}[t]
    \centering
    \includegraphics[width=\linewidth]{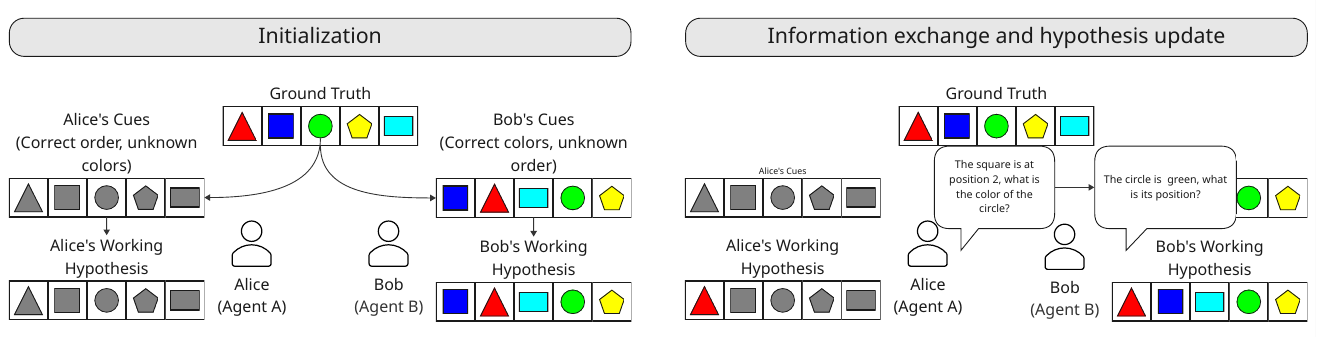}
    \caption{Overview of the puzzle: the ground truth is first created, then each agents' individual partial views are generated and shared with them as clues. The working hypothesis starts as a copy of the clues. Then, in a turn-based interaction, the agents, \agenta and \agentb, can send each other messages and update their working hypothesis until their hypotheses match the ground truth.}
    \label{figure:puzzle_example}
\end{figure}

\papername is a two-agent asymmetric puzzle-solving environment (We provide an overview in \Cref{figure:puzzle_example}) where each participant is given partial information and collaboration is required to reach the solution.

\paragraph{Puzzle setup:} Two agents, \agenta and \agentb, are given complementary views of a position-shape-color matching puzzle. \agenta is given a puzzle view with correct positions and shapes, but unknown colors, while \agentb is given a puzzle view with correct shapes and colors, but unknown positions. The puzzle state is controlled externally; both agents can communicate with each other and provide a list of actions to apply to their \textit{working hypothesis} (i.e., their current view of the puzzle).
\paragraph{Game loop:} A puzzle is generated and the \textit{clues} (initial views) of \agenta and \agentb are separated. Multiple turns take place, where a turn is: \agenta is prompted with the puzzle instruction, cues, working hypothesis, message history (both Alice's and Bob's messages), feedback from the previous turn, and the structure of the format they must respond in. \agenta generates its output, which contains a message for \agentb and the list of actions to take on its puzzle. \agentb follows the same procedure with the latest message from \agenta. The actions are applied to the working hypotheses, which are compared against the ground truth to provide feedback for the next turn. This process continues until the puzzle is solved.
The agents must provide a formatted JSON at the end of their answer, which contains a list of formatted actions and a message that will be shared with the other agent.

\paragraph{Feedback modes:} Feedback is provided to both \agenta and \agentb as part of their input prompts. The feedback can be 1) \textit{No feedback}: no feedback is provided to the agents 2) \textit{Own}: each agent is told whether its part of the puzzle is solved, 3) \textit{Own detailed}: each agent is told whether its current puzzle is solved and which positions are wrong 4) \textit{Joint}: the agents are told whether the puzzle is solved (both of them need to have solved it) 4) \textit{Both}: the agents are told whether their and the other agent's parts of the  puzzle are solved 5) \textit{Both detailed}: the agents get the equivalent of \textit{Own detailed} but for both agents.

\paragraph{Difficulty levels:} The search space for puzzles with $N$ positions and two attributes per position is $(N!)^{2}$ \cite{liuAutonomousAgentsCollaborative2024}. Without communication, Alice's and Bob's problems allow for $N!$ possible permutations; nonetheless, given full information, the puzzle can be solved in linear time $O(N)$, where Alice and Bob only need to map the position-shape-color to their current working hypothesis.

\section{Experimental Setup}

\subsection{Models}
We evaluated the \papername environment on a number of LLMs from various vendors: OpenAI (GPT-3.5-turbo, GPT-4o \cite{openaiGPT4oSystemCard2024}, GPT-5, OSS-120B \cite{openaiGptoss120bGptoss20bModel2025}), Meta (Llama  3.2-11B \cite{grattafioriLlama3Herd2024}), and Anthropic (Claude-3.5, Claude-4.0).
This selection captures a range of reasoning abilities, response tendencies, and cost profiles.
Models differ in training scale, safety alignment, and conversational style, allowing us to examine how choice of LLM affects multi-agent communication and problem-solving.

\subsection{Evaluation Metrics}
For each of our experiments, we set the maximum number of turns to be twice the number of elements in the puzzle; thus, we cut off a 5-piece game after 10 turns.
Note that with full information sharing, a puzzle of any size can be solved in 2 turns. Assuming a single piece of information is exchanged each turn, the maximum number of turns still provides margin for error and correction.

\paragraph{Success Percentage:} Given a set of puzzles to solve -here simulated via seeds- we compute the percentage of these puzzles that are solved within the maximum number of turns.

\paragraph{Average number of actions per position :} For each position, we count the number of times the position is modified. This allows estimation of how many times an agent overwrites its working hypothesis, an indication of whether its decisions are error corrections or just random guesses.

\section{Results} \label{section:results}
\subsection{Can the agent communicate and solve the puzzle?}
We observed that while the task could be successfully solved by a single agent when provided with all of the information, the two agent problem is challenging for most models. (\Cref{table:main-completition})
We note that GPT-5 and Claude-4.0 can consistently solve the task for size 5, achieving 100\% completion regardless of the type of feedback, whereas other models benefit from different feedback strategies.

\subsection{What is the impact of feedback?}
We evaluated different forms of feedback for puzzles with five elements and observed that providing individual feedback on each agent's own working hypothesis increased the completion rate. Detailed feedback led to the most improvement, for instance, on GPT-4o: from $43.3\%$ to $63.3\%$ completion.
On the other hand, providing detailed information about the other agent's working hypothesis on top of their own hurt performance; this can be attributed to information overload with lack of context (Alice does not see Bob's working hypothesis, yet Alice is told which positions of Bob's hypothesis are incorrect).

\subsection{Do agents over-correct their working hypothesis?}
In the optimal solution, \agenta should only modify each position once and \agentb at most once (the working hypothesis could already have correct entries).
Both agents would nominally gather the information they need about a position, update the position, and refrain from modifying it further.
We observe that GPT-5 and Claude-4.0 are nearly optimal as they usually need only two turns of communication and their agents share all the information they have with one another. The other models tend to exchange one piece of information at a time, and models such as GPT-3.5-turbo and Llama 3.2-11B tend to ignore each other's messages. (See appendix \Cref{figure:actions_per_position}).

\subsection{What is the impact of the puzzle size?}
We further evaluate the performance of the different models on puzzles of size 3, 10, and 20, using joint feedback. We adjust the maximum number of turns to 6, 20, and 40, respectively.
We observe that as the puzzle complexity increases, the performance decreases, similarly to \cite{linZebraLogicScalingLimits2025}.

\begin{table}[]
    \centering
    \begin{tabular}{lrrrrrr}
    \toprule
    & \multicolumn{3}{c}{Own Feedback} & \multicolumn{3}{c}{Joint Feedback} \\ 
    Feedback Mode & No feedback & Own & Own Detailed & Joint & Both  & Both detailed \\ 
    Model &  &  &  &  &  & \\
    \midrule
    GPT-5 & 100.0 & 100.0 & 100.0 & 100.0 & 100.0  & 100.0  \\
    Claude-4.0 & 100.0 & 100.0 & 100.0 & 100.0 & 100.0  & 100.0  \\
    OSS-120B & 53.3 & 93.3 & 100.0 & 90.0 & 90.0 & 96.7 \\
    GPT-4o & 43.3 & 46.7 & 63.3 & 40.0 & 80.0  & 56.7  \\
    Claude-3.5 & 66.7 & 73.3 & 86.7 & 73.3 & 83.3  & 36.7  \\
    GPT-3.5-turbo & 0.0 & 0.0 & 0.0 & 0.0 & 0.0 & 0.0 \\
    Llama 3.2-11B & 0.0 & 0.0 & 0.0 & 0.0 & 0.0 & 0.0 \\
    \bottomrule
\end{tabular}

    \caption{(Higher is better) Percentage of 5-pieces puzzles solved over 30 seeds. Temperature 0.0. Providing feedback increases the completion percentage, while additionally providing detailed information about the other agent's working hypothesis can hurt performance. (Tables with Wilson 95\% Confidence Intervals are provided in Appendix \Cref{appendix:table:wci_own} and \Cref{appendix:table:wci_joint})}
    \label{table:main-completition}
\end{table}

\begin{table}[]
    \centering
    \begin{tabular}{lrrrr}
    \toprule
    & \multicolumn{4}{c}{Both feedback} \\ 
    Puzzle size & 3 & 5 & 10 & 20 \\ 
    Model &  &  &  & \\
    \midrule
    GPT-5 & 100.0 & 100.0 & 100.0 & 100.0   \\
    Claude-4.0 & 100.0  & 100.0 & 100.0 & 100.0 \\
    OSS-120B & 83.3 & 90.0 & 90.0 & 93.3  \\
    GPT-4o & 60.0 & 80.0 & 60.0 & 16.7  \\
    Claude-3.5 & 50.0  & 83.3 & 56.7 & 16.7   \\
    GPT-3.5-turbo & 0.0 & 0.0 & 0.0 & 0.0 \\
    Llama 3.2-11B & 0.0 & 0.0 & 0.0 & 0.0  \\
    \bottomrule
\end{tabular}

    \caption{(Higher is better) Success rate for different size puzzles across 30 seeds, with temperature 0.0, and each agent receiving both agent's solved status. As complexity increases, GPT-4o and Claude-3.5 decrease in performance. The lower performance by some models on the 3-pieces puzzle can be explained by miscommunication, leading to wasted turns under a tighter turn constraint.}
    \label{table:main-size}
\end{table}

\section{Conclusion} \label{section:conclusion}
We presented Asympuzl, an evaluation testbed for multi-turn cooperative play between LLM agents under information asymmetry.
We demonstrated that strong models (e.g., GPT-5 and Claude-4.0) can reliably converge with each agent sharing all of its information as agents ideally would.
In contrast, other models struggle with miscommunication or repeated corrections.
Our results show that feedback matters: simple self-feedback improves performance, but detailed joint feedback can confuse agents.
These findings underscore the importance of carefully designing communication and evaluation protocols in multi-agent LLM systems.

Looking ahead, AsymPuzl can serve as a foundation for more complex studies, such as introducing noisy or ambiguous views, restricting communication bandwidth, or scaling to three or more agents.
We hope this testbed will help investigate coordination strategies and contribute to solving the broader challenges of enabling LLMs to collaborate effectively in real-world, multi-turn settings.

\begin{ack}
This research was funded by the Defense Advanced Research Projects Agency (DARPA), under contract W912CG23C0031.
\end{ack}

\printbibliography

\clearpage
\appendix
\section*{Outline} \label{section:appendix}

\begin{itemize}
    \item[]\Cref*{appendix:limitations} \nameref{appendix:limitations}
    \item[]\Cref*{appendix:metrics} \nameref{appendix:metrics}
    \item[]\Cref*{appendix:single} \nameref{appendix:single}
    \item[]\Cref*{appendix:confidence} \nameref{appendix:confidence}
    \item[]\Cref*{appendix:implementation} \nameref{appendix:implementation}
    \item[]\Cref*{appendix:communication} \nameref{appendix:communication}
    \item[]\Cref*{appendix:prompt} \nameref{appendix:prompt}
\end{itemize}

\section{Limitations} \label{appendix:limitations}
\paragraph{Instruction and puzzle injection:} We adopt a design where the environment re-injects all relevant state into each turn, namely, the agents are provided with the set of instructions, output format requirements, initial cues, current working hypotheses, and feedback about the current working hypotheses.
This allows isolating the role of the communication strategy, and avoiding the agent losing track of its original task.
While this differs from fully autonomous agent memory, it still constitutes a multi-turn interaction, as agents must iteratively exchange complementary information and solve the task through sequential coordination.
We leave extensions toward persistent agent state as future work.

\paragraph{Original cue injection:} At each turn, we provide the agents with their original cues.
We chose this design so that the agents always have a way to recover their original information. Furthermore, this allows the agent to compare its working hypothesis to its original cues, keeping track of its progress.
To translate this to a real-world setting, it would be similar to having the agent query a database and caching the information in a read-only entry so that the agent cannot overwrite the data it is accessing.

\paragraph{Puzzle complexity:} We leave for future work the analysis of noisy and ambiguous interactions where, for instance, \agenta is given more shapes than required while \agentb has the correct number of colors to shape pairs, requiring them to determine the relevant information. This ambiguity can be applied in the opposite direction as well, where \agentb would see more color-shape associations than there are valid shapes in the ground truth.
Another extension is the evaluation of constraints on communication, or the number of operations per turn, and the addition of an internal scratch pad carried across turns by the agents.

\section{Evaluation Metrics} \label{appendix:metrics}

\begin{figure}[ht]
  \centering
  \begin{subfigure}{0.45\textwidth}
    \includegraphics[width=\linewidth]{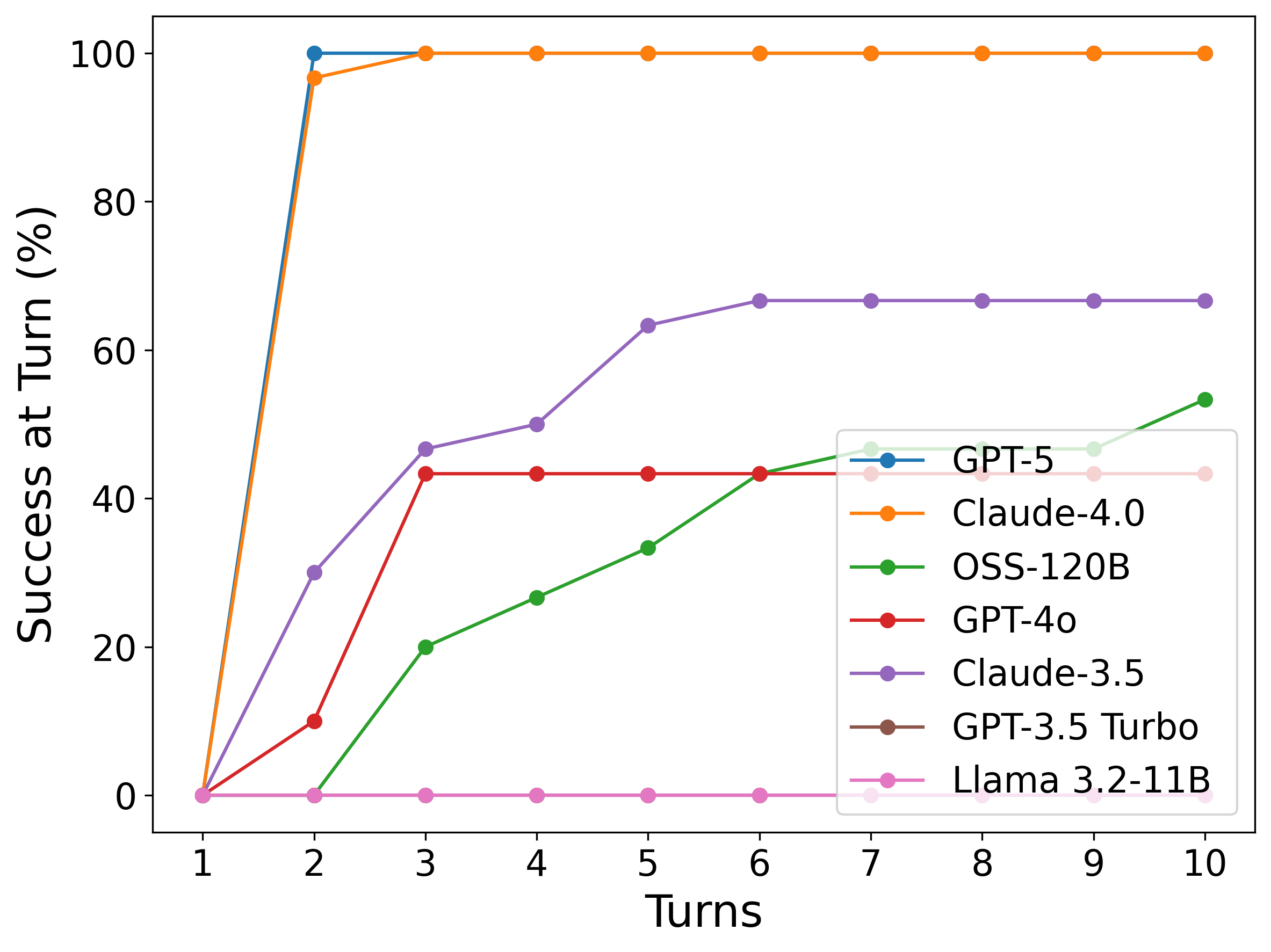}
    \caption{Success rate at different turns when the agents are not given any feedback.}
    \label{figure:first}
  \end{subfigure}
  \hfill
  \begin{subfigure}{0.45\textwidth}
    \includegraphics[width=\linewidth]{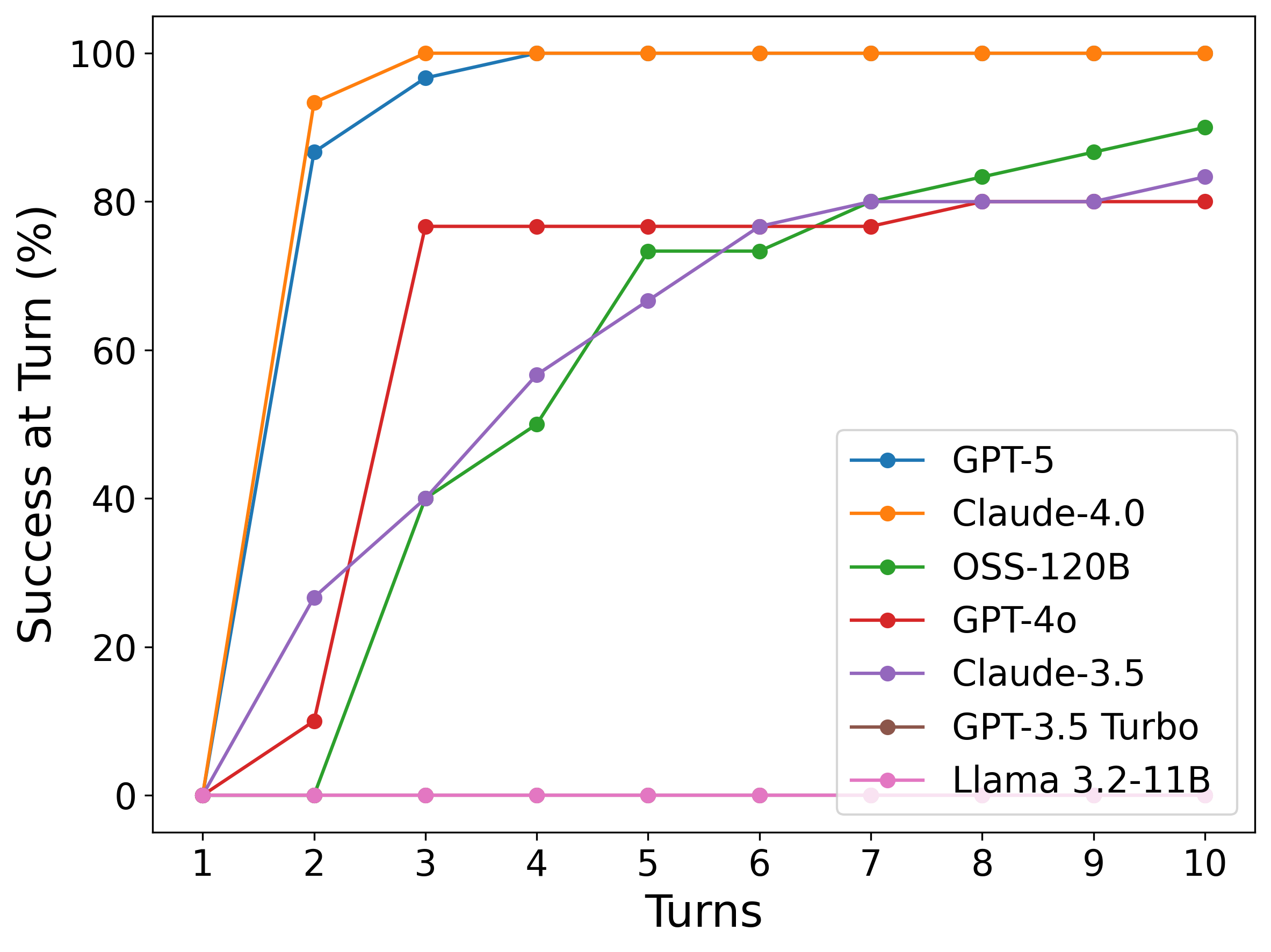}
    \caption{Success rate at different turns when both agents get joint feedback.}
    \label{figure:second}
  \end{subfigure}
  \caption{Comparison of the success rate of each LLM model over time (turns) for 5-piece puzzles with No feedback or Both feedback. Providing feedback about both sides of the puzzle increases success rate.}
  \label{figure:success_rate_at_k}
\end{figure}

\paragraph{Success rate at Turn K:} We gather across experimental seeds the number of turns taken to solve each puzzle. This additionally allows us to derive the number of puzzles solved within each given number of turns. We compared two feedback modes, No Feedback and the Joint feedback where both agent see each other's completion status (\Cref{figure:success_rate_at_k}). This experiment showed that the joint feedback increased the success rate and that models such as GPT-5, Claude 4.0, and GPT-4o tend to solve the puzzles in fewer turns.

\paragraph{Number of tokens:} We collected the average number of tokens that each agent output per turn and the number of these tokens dedicated to the message sent to the other agent (\Cref{table:num_tokens}). We observed that for most models \agentb uses more token, this could be explained by the need to associate both shape and colors in its message while \agenta can enumerate the shapes. To compare the different model we use the same tokenizer from tiktoken across models' outputs.

\begin{figure}[h]
    \centering
    \includegraphics[width=\linewidth]{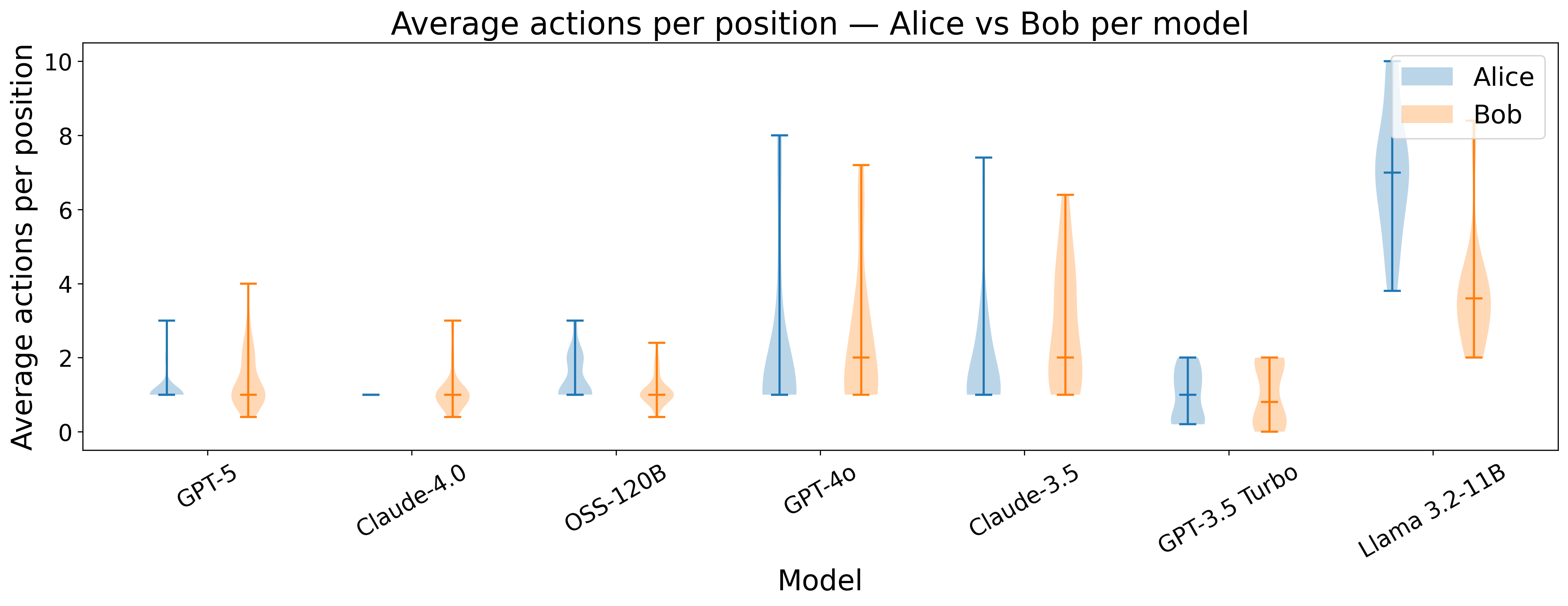}
    \caption{(Lower is better) Average number of actions per position. GPT-5 and Claude-4.0 are close to optimal on average with few positional modifications. Meanwhile, GPT-3.5-turbo tends not to modify positions despite the puzzle being unsolved, and Llama 3.2-11B tends to modify positions more than 4 times on average.}
    \label{figure:actions_per_position}
\end{figure}

\begin{table}[]
    \centering
    \begin{tabular}{lcccccc}
    \toprule
    & \multicolumn{3}{c}{Agent A} & \multicolumn{3}{c}{Agent B} \\ 
    \#Tokens & Output & Message & $\frac{\text{Message}}{\text{Output}} \times 100$ & Output &  Message & $\frac{\text{Message}}{\text{Output}} \times 100$ \\
    Model & & & & & & \\
    \midrule
    GPT-5 & 155.3 & 59.4 & 38.2 & 185.8 & 69.5 & 37.4 \\
    Claude-4.0 & 225.6 & 68.7 & 30.5 & 311.9 & 70.2 & 22.5 \\
    OSS-120B & 94.3 & 26.4 & 28.0 & 77.8 & 23.8 & 30.7 \\
    GPT-4o&114.3 & 27.3 & 23.9 & 137.8 & 33.2 & 24.1 \\
    Claude-3.5 & 188.6 & 32.2 & 17.1 & 223.7 & 39.1 & 17.5 \\
    GPT-3.5-turbo &40.6 & 8.4 & 20.8 & 41.4 & 8.2 & 19.8 \\
    Llama 3.2-11B &94.5 & 8.6 & 9.1 & 62.9 & 8.4 & 13.4 \\
    \bottomrule
\end{tabular}

    \caption{Comparison of the average number of tokens across agents when both agents are provided with feedback about each other's completion status. Claude models tend to be more verbose, and for the GPT-5 model, close to 40\% of the generated output is dedicated to the message sent to the other agent.}
    \label{table:num_tokens}
\end{table}

\section{Solving as a Single Agent with Full Information} \label{appendix:single}
To verify that the LLMs could solve the problem with complete information, we conducted multiple experiments of a single agent solving the puzzle across 10 seeds.
(\Cref{table:single}).

We used this to evaluate the problem presentation and ensure that potential difficulties arising from the two-agent setup would stem from multi-agent interactions rather than single-agent issues. Most LLM agents achieved a completion rate of 100\% across different puzzle sizes. The agents are only given a single attempt without feedback.

\begin{table}[h]
    \centering
    \begin{tabular}{lccc}
\toprule
& \multicolumn{3}{c}{Success \% ($\uparrow$)} \\
Puzzle Size & 5 & 10 & 20 \\
Model &  &   & \\
\midrule
GPT-5 & 100.0 & 100.0 & 100.0 \\
Claude-4.0 & 100.0 & 100.0 & 100.0 \\
OSS-120B & 100.0 & 100.0 & 100.0 \\
GPT-4o & 100.0 & 100.0 & 100.0 \\
Claude-3.5 & 100.0 & 86.7 & 93.3\\
GPT-3.5-turbo & 100.0 & 100.0 & 96.7 \\
Llama 3.2-11B & 80.0 & 96.7 & 96.7 \\
\bottomrule
\end{tabular}

    \caption{Evaluation of a single agent with full information solving puzzles of different sizes.}
    \label{table:single}
\end{table}

\section{Tables with 95\% Confidence Intervals} \label{appendix:confidence}
We provide the 95\% confidence intervals for the values in \Cref{table:main-completition}

\begin{table}[h]
    \centering
    \begin{tabular}{lccc}
    \toprule
    Feedback Mode & No feedback (CI) & Own (CI) & Own Detailed (CI) \\
    Model &  &  &  \\
    \midrule
    GPT-5 & $100.0_{(88.6 - 100.0)}$ &  $100.0_{(88.6 - 100.0)}$ & $100.0_{(88.6 - 100.0)}$  \\
    Claude-4.0 & $100.0_{(88.65,100.00)}$ &  $100.0_{(88.65,100.00)}$ & $100.0_{(88.6 - 100.0)}$  \\
    OSS-120B & $53.3_{(36.1 - 69.8)}$ & $93.3_{(78.7 - 98.2)}$ & $100.0_{(88.6 - 100.0)}$  \\
    GPT-4o & $43.3_{(27.4 - 60.8)}$ & $46.7_{(30.2 - 63.9)}$ & $63.3_{(45.5 - 78.1)}$  \\
    Claude-3.5 & $66.7_{(48.8 - 80.8)}$ & $73.3_{(55.6 - 85.8)}$ & $86.7_{(70.3 - 94.7)}$ \\
    GPT-3.5-turbo & $0.0_{(0.0 - 11.4)}$ & $0.0_{(0.0 - 11.4)}$ & $0.0_{(0.0 - 11.4)}$ \\
    Llama 3.2-11B & $0.0_{(0.0 - 11.4)}$ & $0.0_{(0.0 - 11.4)}$ & $0.0_{(0.0 - 11.4)}$ \\
    \bottomrule
\end{tabular}

    \caption{(Higher is better) Percentage of 5-element puzzles solved over 30 seeds, temperature 0.0, with Wilson 95\% Confidence Intervals, for the own feedback.}
    \label{appendix:table:wci_own}
\end{table}

\begin{table}[h]
    \centering
    \begin{tabular}{lccc}
    \toprule
    Feedback Mode & Joint (CI) & Both (CI) & Both detailed (CI) \\ 
    Model &  &  &  \\
    \midrule
    GPT-5 & $100.0_{(88.6 - 100.0)}$ & $100.0_{(88.6 - 100.0)}$ & $100.0_{(88.6 - 100.0)}$ \\
    Claude-4.0 & $100.0_{(88.6 - 100.0)}$ & $100.0_{(88.6 - 100.0)}$ & $100.0_{(88.6 - 100.0)}$  \\
    OSS-120B &  $90.0_{(74.4 - 96.5)}$ & $90.0_{(74.4 - 96.5)}$ & $96.7_{(83.3 - 99.4)}$  \\
    GPT-4o & $40.0_{(24.6 - 57.7)}$ & $80.0_{(62.7 - 90.5)}$ & $56.7_{(39.2 - 72.6)}$  \\
    Claude-3.5 & $73.3_{(55.6 - 85.8)}$ & $83.3_{(66.4 - 92.7)}$ & $36.7_{(21.9 - 54.5)}$  \\
    GPT-3.5-turbo & $0.0_{(0.0 - 11.4)}$ & $0.0_{(0.0 - 11.4)}$ & $0.0_{(0.0 - 11.4)}$  \\
    Llama 3.2-11B & $0.0_{(0.0 - 11.4)}$ & $0.0_{(0.0 - 11.4)}$ & $0.0_{(0.0 - 11.4)}$ \\
    \bottomrule
\end{tabular}

    \caption{(Higher is better) Percentage of 5-element puzzles solved over 30 seeds, temperature 0.0, with Wilson 95\% Confidence Intervals, for the joint feedback.}
    \label{appendix:table:wci_joint}
\end{table}

\section{Implementation Details and Hyper-parameters} \label{appendix:implementation}
\subsection{Implementation}
We first build the puzzle, determine the original clues for both agent, and use an environment to monitor and provide the current working hypothesis of both agents. This allows the agent to focus on solving the tasking and on communicating information rather than trying to represent the puzzle. We use LangChain to query the different agents, and for the open-source models we host them using vLLM~\cite{kwonEfficientMemoryManagement2023}.

\subsection{Hyper-parameters}
For all models, we use a maximum of $4,096$ output tokens, set temperature to $0.0$, and repeat experiments across $30$ seeds, where the seed controls the puzzle generation and initial clues.
We further provide a history length of $1$ to the agents, namely they see their previous message and the latest message from the other agent.

\section{Communication Issues} \label{appendix:communication}
We provide examples of communication between \agenta and \agentb.
In our initial prompt design, we noticed that some agents were not sharing information.

We hypothesized that they were assuming the other agent would see both their message and their actions. We detailed the prompt further, indicating that the other agent would only see the "message" they sent, and the action would not be shared.

We provide excerpts of the conversations for 3 cases \nameref{appendix:communication:success}, \nameref{appendix:communication:no_coop} and \nameref{appendix:communication:failed_with_confidence}

\subsection{Successful collaboration} \label{appendix:communication:success}
\begin{figure}[h]
    \centering
    \includegraphics[width=\linewidth]{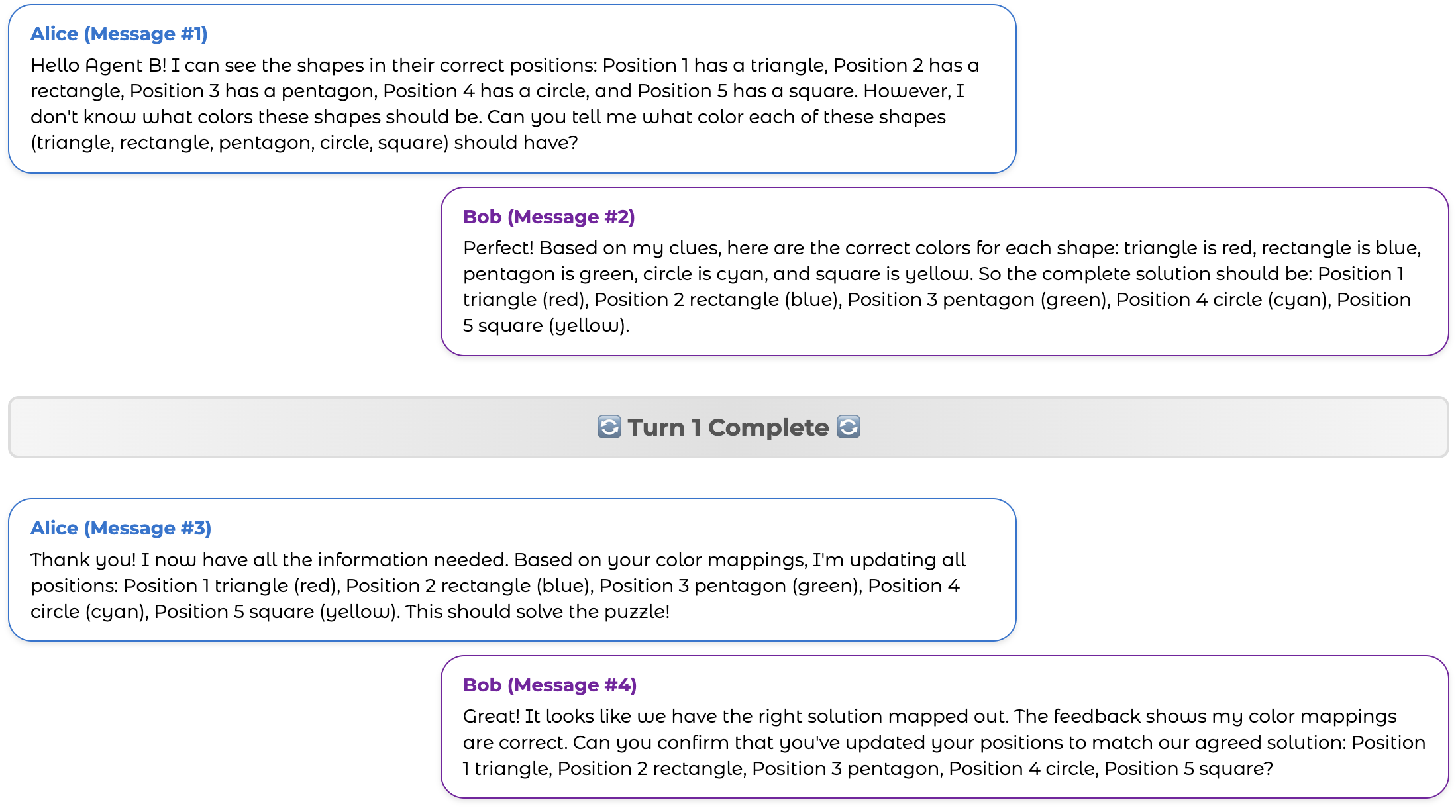}
    \caption{Example of successful completion using Claude 4.0. By the end of the second turn, the puzzle is solved, as both agents shared all of their information and cooperated.} \label{figure:communication_success}
\end{figure}

We provide an example of successful communication where the two agents manage to share the most information and solve the puzzle in two turns. (\Cref{figure:communication_success})

\subsection{No cooperation} \label{appendix:communication:no_coop}
\begin{figure}[h]
    \centering
    \includegraphics[width=\linewidth]{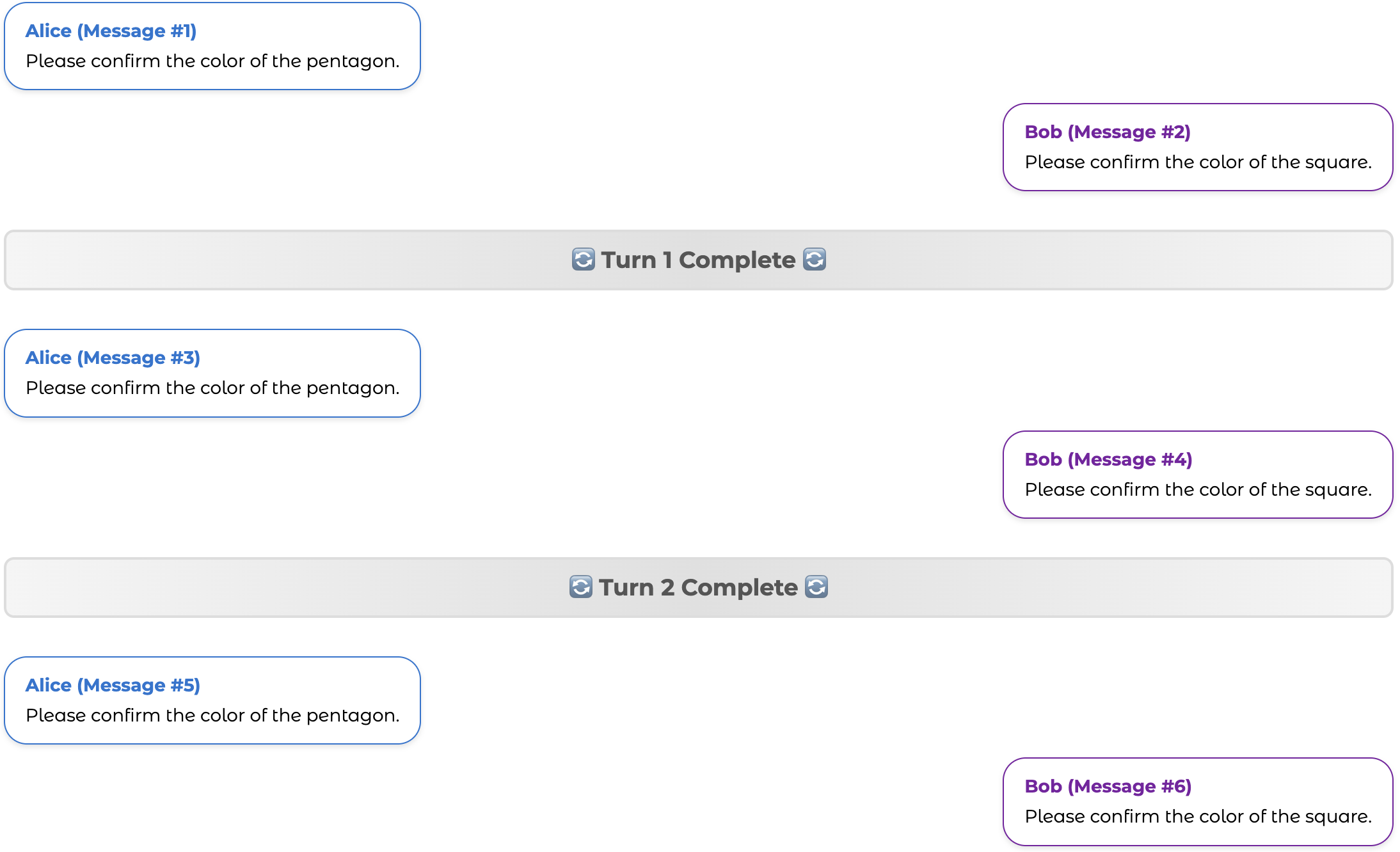}
    \caption{Example of lack of cooperation using Llama 3.2-11B. Both agents ignore one another's messages.}
    \label{figure:communication_no_coop}
\end{figure}

We provide an example where both agents ignore one another's messages and keep repeating their previous message instead. (\Cref{figure:communication_no_coop})

\subsection{Miscommunication} \label{appendix:communication:failed_with_confidence}
\begin{figure}[h]
    \centering
    \includegraphics[width=\linewidth]{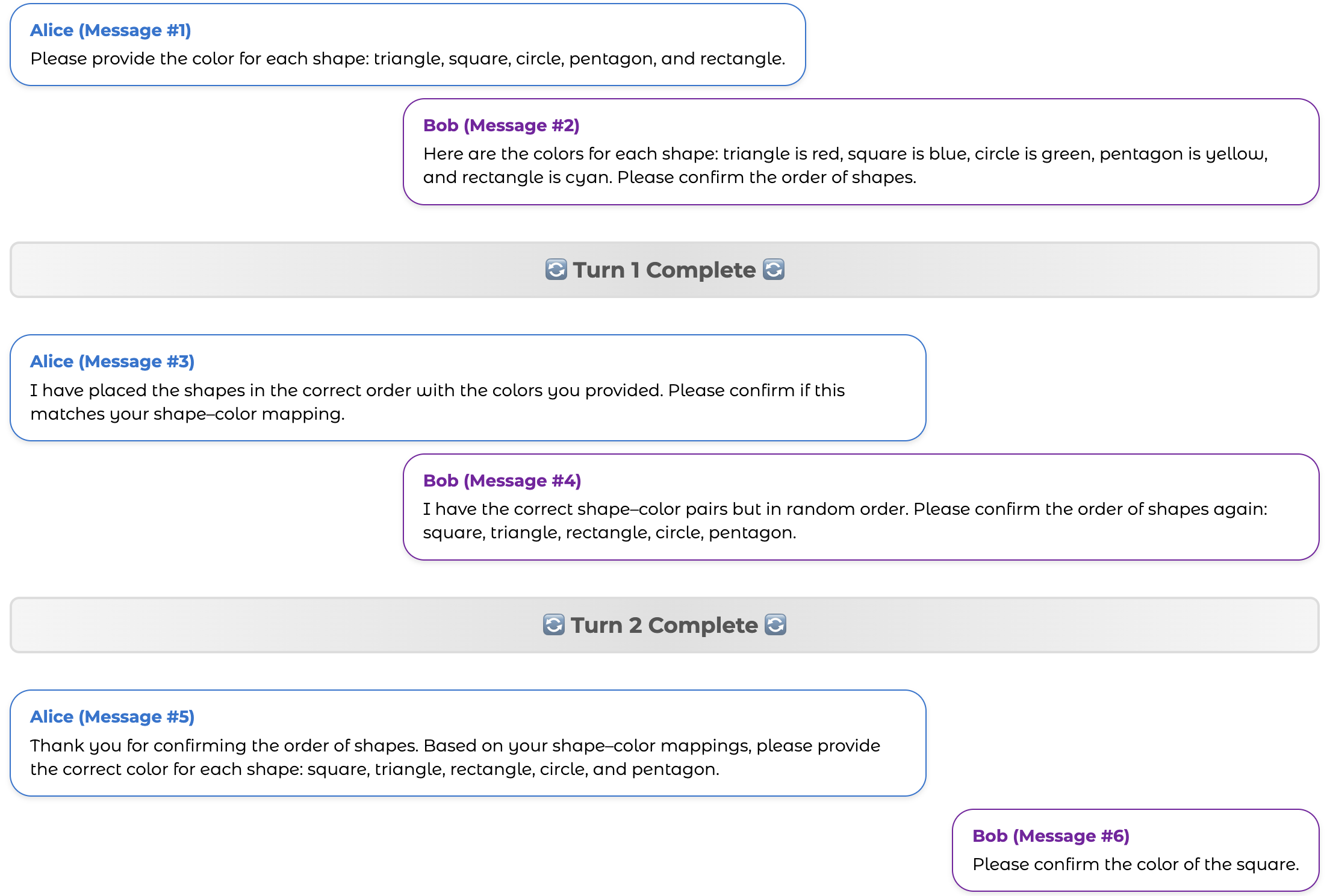}
    \caption{An example of cooperation with miscommunication using GPT-4o. \agenta provides the correct order and \agentb provides the color information but dismisses the position. \agenta adapts the colors but assumes that \agentb can see its view and asks it to confirm that the mapping matches. \agentb provides a new order that does not match the original order provided by \agenta.}
    \label{figure:communication_miscommunication}
\end{figure}

We provide an example where one agent makes assumptions on what the other agent can see, and the other agent provides information that it has no guarantee about (\Cref{figure:communication_miscommunication}). This leads to the agents getting sidetracked and losing multiple turns before potentially recovering or running out of turns.

\clearpage
\newpage
\section{Example of prompt} \label{appendix:prompt}
We provide an example of the prompt given to \agentb.
The part before the \#Puzzle State is given as a system prompt.

\begin{tcolorbox}[title=Example of prompt for \agentb,
                  colback=gray!10,
                  colframe=black,
                  breakable,
                  parskip=1ex]
You are Agent B. \\
Your goal is to solve a symbolic puzzle with 5 positions. \\
Each position must have the correct (shape, color) pair. \\
- You see the correct shape–color mappings, but in random order. \\
- Agent A knows the correct order of shapes but not the colors.\\
Communicate with Agent A and update your hypothesis until the puzzle is solved. \\

\# Puzzle State \\
Initial clues: \\
Position 1: the square is blue. Position 2: the triangle is red. Position 3: the rectangle is cyan. Position 4: the circle is green. Position 5: the pentagon is yellow. \\

Your current working hypothesis: \\
Position 1: the square is blue. Position 2: the triangle is red. Position 3: the rectangle is cyan. Position 4: the circle is green. Position 5: the pentagon is yellow. \\

Feedback: \\
Agent A's part of the puzzle is solved and Agent B's part of the puzzle is unsolved. \\

Recent conversation: \\
B: I have the correct shape–color pairs but in random order. Please confirm the order of shapes again: square, triangle, rectangle, circle, pentagon. \\
A: Thank you for confirming the order of shapes. Based on your shape–color mappings, please provide the correct color for each shape: square, triangle, rectangle, circle, and pentagon. \\

Output format: \\
Each action must follow this format:  

\begin{verbatim}
{"replace": <position>, "by": {"shape": <shape>, "color": <color>}}
\end{verbatim}

Note: <position> uses 1-based indexing (position 1 is the first item, position 2 is the second item, etc.) \\
Your answer MUST END WITH a **valid JSON object** and include the following fields: \\
- "message": What you want to tell and ask to the other agent (the only thing the other agent will receive). \\
- "actions": A list of actions to take (the other agent will not see your actions). \\

Example:
\begin{verbatim}
```json
{
  "message": "Please confirm the color of the circle.",
  "actions": [
    {
      "replace": 1,
      "by": {
        "shape": "circle",
        "color": "red"
      }
    }
  ]
}
```
\end{verbatim}
\end{tcolorbox}

\end{document}